\documentclass[12pt]{iopart}
\usepackage{hyperref} 
\usepackage{doi} 
\usepackage{iopams}
\usepackage{graphicx}
\usepackage{xcolor}

\makeatletter
\newcommand*{\rom}[1]{\expandafter\@slowromancap\romannumeral #1@}
\makeatother

\begin{document}
\title[Magnetization induced skyrmion dynamics]{Magnetization induced skyrmion dynamics of a spin-orbit-coupled spinor condensate under sinusoidally varying magnetic field}

\author{Arpana Saboo$^1$, Soumyadeep Halder$^1$, Mithun Thudiyangal$^2$, Sonjoy Majumder$^1$}

\address{$^1$ Department of Physics, Indian Institute of Technology Kharagpur, Kharagpur, West Bengal 721302, India}
\address{$^2$ Department of Physics and Electronics, Christ University, Bengaluru, Karnataka 560029, India}

\ead{\mailto{arpana.saboo@gmail.com}, \mailto{sonjoym@phy.iitkgp.ac.in}}

\begin{abstract}
We explore the spin texture dynamics of a harmonically trapped spin-1 Bose-Einstein condensate with Rashba spin-orbit coupling and ferromagnetic spin-exchange interactions under a sinusoidally varying magnetic field along the $x$-direction. This interplay yields an intrinsic spin texture in the ground state, forming a linear chain of alternating skyrmions at the saddle points of the magnetic field. Our study analyzes the spin-mixing dynamics for both a freely evolving and a controlled longitudinal magnetization. The spin-1 system exhibits the Einstein-de Haas effect for the first case, for which an exchange between the total orbital angular momentum and the spin angular momentum is observed, resulting in minimal oscillations about the initial position of the skyrmion chain. However, for the fixed magnetization dynamics, the skyrmion chain exhibits ample angular oscillations about the equilibrium position, with the temporary formation of new skyrmions to facilitate the oscillatory motion. For the case of fixed magnetization, this contrast now stems from the exchange between the canonical and spin-dependent contribution to the orbital angular momentum. The variation in canonical angular momentum is linked to the angular oscillations, while the spin-dependent angular momentum accounts for the creation or annihilation of skyrmions. We confirm the presence of scissors mode excitations in the spin texture due to the angular skyrmion oscillations. 
\end{abstract}
\vspace{2pc}
\noindent{\it Keywords}: spinor condensates, spin-1 BECs, spin-orbit coupling, skyrmions, spin texture dynamics, spin-spin correlation

\submitto{\NJP}
\maketitle

\section{Introduction}
\label{sec:intro}
Spinor Bose gases \cite{hall_dynamics_1998, hall_measurements_1998, matthews_dynamical_1998,ho_spinor_1998, ohmi_bose-einstein_1998} form a family of quantum fluids with non-trivial spin internal degrees of freedom, manifesting the interplay of magnetism \cite{koashi_exact_2000} and superfluidity \cite{Flayac_superfluidity_2013}, both of which involves quantum phase coherence \cite{eto_dissipation_2019}, long-range order \cite{Kroeze_spinor_2018}, and symmetry breaking \cite{Scherer_spontaneous_2013, hoshi_symmetry_2010}. These degenerate fluids are interesting in their own right as they could be investigated for the system properties and the interaction between atomic states, shedding light on a range of topics such as quantum phase transitions \cite{Sachdev_1999}, non-equilibrium quantum dynamics \cite{huh_universality_2024, huh2024skyrmionspintexturequantum} and instabilities \cite{jose_patterns_2023, saboo_rayleigh_2023},  the role of symmetry and topology in quantum-ordered materials \cite{kumar_topological_2021,ozawa_topological_2019}, topological defects \cite{saboo_sinusoidal_2024,mithun_existence_2022}, spin dynamics \cite{eto_spinor_2018}, etc. 

The experimental realization of spin-orbit coupling (SOC) \cite{lin_spinorbit-coupled_2011,wang_spinorbit_2010, Hui_soc_2012, wang_spin-orbit_2012,anderson_synthetic_2012,wu_realization_2016, huang_experimental_2016,campbell_realistic_2011} in neutral atoms has broadened the impact of ultracold-atom research opening a new avenue for spinor quantum gases. Spin-orbit (SO) coupled spin-1 BECs are particularly interesting to explore as they exhibit certain exotic quantum states with engrossing topological structure in their order parameter which includes vortex lattice \cite{wang_vortex_2017, wang_vortex_2020, liu_vortex_2013}, half-quantum vortex \cite{Hui_soc_2012, ramachandhran_half-quantum_2012, gautam_fractional_2016}, supersolid stripe phase \cite{li_stripe_2017, zhao_magnetic_2020, geier_dynamics_2023}, solitons \cite{mithun_stationary_2024}, skyrmions \cite{kawakami_stable_2012, luo_three-dimensional_2019, su_crystallized_2012}, among others. The degenerate spinor condensates are, therefore, an enticing bridge between the physics of ultracold atoms and solid states.

The SO-coupled spinor gas offers a unique opportunity to study topological excitations with spin textures in a controlled environment. One of the most fascinating outcomes of introducing SOC in spin-1 BECs is the emergence of skyrmion spin textures \cite{Ueda_2014, zhang_manipulation_2018}. Skyrmions are topologically protected spin configurations that resemble spin vortices \cite{kato_twisted_2016}, stabilized by the interplay of SOC and atomic interactions.  Investigating skyrmion textures enhances our understanding of topological phases of matter and quantum vortices \cite{choi_observation_2012}, with potential applications in quantum computing \cite{Smith_crossing_2022}, spintronics \cite{marrows_perspective_2021}, Spin Hall effect \cite{Chen_Skyrmion_2019, kimbell_challenges_2022}, and advanced materials \cite{Rana_Skyrmions_2023}. The study of skyrmions in SO-coupled spin-1 BECs underscores the profound implications of topological and spin-orbit phenomena in quantum fluids, paving the way for significant advancements in both theoretical and applied physics \cite{liu_spin_orbit_2012, Liu_bimeron_2024, zhu_spin_2020,choi_imprinting_2012, liu_composite_2020}. Moreover, the interaction of the SO-coupled BECs with the external magnetic field yields the local spin texture of the system \cite{anderson_magnetically_2013, yang_topological_2022, yang_dynamics_2022}. Recent studies indicate that in-plane varying magnetic fields can be used to synthetically create non-Abelian gauge fields, which may lead to exotic phenomena in spin-1 BECs such as skyrmion lattice formation as discussed in \cite{saboo_sinusoidal_2024}.

In the present work, we investigate the dynamics of a spin-orbit-coupled spin-1 Bose-Einstein condensate confined in a quasi-2D harmonic trap and subjected to a sinusoidally varying magnetic field along the $x$-direction. The study focuses on the system's dynamical response, highlighting how the magnetization of the prepared state influences angular momentum transfer mechanisms. Freely evolving magnetization showcases the Einstein-de Haas effect, while fixed magnetization drives orbital angular momentum exchange between canonical and spin-orbit dependent contributions, leading to distinct skyrmion chain oscillations. We further investigate the spin-spin correlations for both cases of spin texture dynamics.

This paper is organized as follows: In section \ref{sec:model}, we introduce the model of our system, casting it to coupled Gross-Pitaevskii (GP) equations. In section \ref{sec:Results}, we present the results obtained by numerically solving the GP equations and investigate the dynamical evolution of the system in section \ref{subsec:3.1} under two cases of [section\ref{subsubsec:free}] freely evolving and [section \ref{subsubsec:fixed}] fixed magnetization. In section \ref{subsection:scissors}, we discuss the scissor mode excitations in the spin texture during the real-time dynamics of the system under both cases. We analyse the spin-spin correlations in section \ref{subsection:correlation} and draw important conclusions from our study in section \ref{sec:conclusion}. We discuss the dependence on magnetization values on the magnetization dynamics in \ref{apx:a} and the dynamical variation of the topological charge in \ref{apx:b} The detailed derivation of the scissors mode in the spin texture is given is \ref{apx:c}.

\section{Model}
\label{sec:model}
 We consider a Quasi-2D spin-orbit coupled $F=1$ spinor condensate of $N$ atoms with mass $M$ confined in an external axisymmetric potential $V(\mathbf{r})$ at zero temperature. In the mean-field theory, the effective Hamiltonian of the system is described by  \cite{kawaguchi_spinor_2012},
\begin{eqnarray}
       \label{eqn:1}
H_{2D} &=& \int d\mathbf{r}  \bigg(\Psi^{\dagger} \left[-\frac{\hbar^2 \nabla^2}{2M} + V(\mathbf{r}) + \nu_{\rm{soc}} + g_F\mu_B \mathbf{B(\mathbf{r}).\mathbf{f}} \right]\Psi \nonumber \\ & &+  \frac{1}{2}g_0 n^2 + \frac{1}{2}g_2|{\mathbf{F}}|^2\bigg)
\end{eqnarray}
where $\Psi =[\Psi_1(\mathbf{r}), \Psi_0(\mathbf{r}), \Psi_{-1}(\mathbf{r})]^T$ with $\mathbf{r}=(x,y)$ is the spin-1 wave function where the three components describe the condensate amplitude in the spin levels $m_F=1,0,-1$, respectively, satisfying the normalization criteria, $\int d\mathbf{r}\Psi^{\dagger}\Psi = N$, with total atomic density $n=n_1 + n_0 + n_{-1} = \sum_{m_F} \Psi^{\dagger}_{m_F} \Psi_{m_F}$ and the spin density given as $ |{\mathbf{F}}| = \Psi^{\dagger}{\mathbf{F}} \Psi$. Here, ${\mathbf{F}} = (F_x, F_y, F_z)$ represents the spin density vector defined as $F_{\alpha} = \Psi^{\dagger}f_{\alpha}\Psi \quad(\alpha = x,y,z)$ with $\mathbf{f}=(f_x, f_y, f_z)$ being the irreducible representation of the $3 \times 3$ Pauli spin matrices and the longitudinal magnetization $m_z= \int d\mathbf{r} F_z(\mathbf{r})$. The spin-independent and spin-exchange interactions are characterized by $g_0$ and $g_2$, respectively, corresponding to channels for which two colliding atoms have total spin $0,2$ respectively. $V(\mathbf{r})=\frac{1}{2}M\omega_{\perp}^2(x^2 + y^2)$ is the Quasi-2D harmonic trap with $\omega_{\perp}$ being the radial trapping frequency and $l=\sqrt{\hbar/M\omega_{\perp}}$ being the harmonic oscillator length. The Rashba SOC interaction \cite{lin_spinorbit-coupled_2011,Hui_soc_2012} yields as $\nu_{\rm{SOC}} = \gamma (f_xp_y - f_yp_x)$ where $\gamma$ is the SOC strength and $(p_x,p_y)$ represents the momentum in the quasi-2D space. $B(\mathbf{r})=(B_0 \sin{x},0)$ is the externally applied magnetic field. Here, the Lande's g-factor $g_F=-\frac{1}{2}$ and $\mu_B$ is the Bohr magneton. In our study, we consider the effective $B_0$ to be negative.

We cast the system in a Quasi-2D form by considering a Gaussian ansatz along the axial direction, thereby integrating it out. The Quasi-2D dimensionless Gross-Pitaevskii equations (GPEs) governing the dynamics of the spin-1 system reads as,

\begin{eqnarray}
    \label{eqn:2}
    {i}{\partial_t{\psi_{\pm 1}}} & = & \big[-\frac{1}{2}\nabla^2 + V + c_0|\psi|^2 
    + c_2\big(\pm|\psi_1|^2 + |\psi_0|^2 \mp |\psi_{-1}|^2\big)\big]\psi_{\pm1} \nonumber \\
    & & + c_2\psi_{\mp 1}^*\psi_0^2 -{i}{\gamma}\big({\partial_y} \pm {i}{\partial_x}\big)\psi_0 
    + B_0\sin{x} \psi_0.
\end{eqnarray}

\begin{eqnarray}
    	\label{eqn:3}
  {i}{\partial_t{\psi_0}} & = & \big[ -\frac{1}{2}\nabla^2 + V + c_0|\psi|^2 + c_2\big(|\psi_1|^2 + |\psi_{{-}1}|^2\big)\big]\psi_{0}  + 2c_2\psi_{1}\psi_0^*\psi_{-1}  \nonumber \\ & &-{i}{\gamma}\big({\partial_y} - {i}{\partial_x}\big)\psi_1 -{i}{\gamma}\big({\partial_y} + {i}{\partial_x}\big)\psi_{-1}  + B_0\sin{x}(\psi_1 +\psi_{-1}) 
  \end{eqnarray}
where $\psi_{m_F}= N^{-1/2}l\Psi_{m_F} \quad (m_F=\pm 1,0)$ denotes the dimensionless $m_F$-th component wave function and $|\psi|^2 = \sum_{m_F} |\psi_{m_F}|^2$ is the total particle density confined in dimensionless external potential $V= (x^2 + y^2)/2$. The contact interaction parameters in the dimensionless form are given as $c_0= 2N\sqrt{2\pi \lambda}(a_0 + 2a_2)/3l$ and $c_2= 2N \sqrt{2\pi \lambda}(a_2 - a_0)/3l$, respectively. In our numerical calculations, the lengths, time, energy (interaction and SOC), and magnetic field are measured in the units of $l, 1/\omega_{\perp}, \hbar\omega_{\perp}$ and $(g_F \mu_B l)/\hbar \omega_{\perp}$, respectively. We use the split-step Crank-Nicolson scheme \cite{crank_nicolson_1947, ANTOINE20132621,muruganandam_2009_fortranprogramstimedependent} to numerically solve the GP eqns (\ref{eqn:2}-\ref{eqn:3}). In our simulations, we have considered $10^4$ number of $\rm{^{87}{Rb}}$ atoms confined in a quasi-2D harmonic trap $V$ with transverse trapping frequency, $\omega_{\perp}=2\pi \times 20$ Hz with trap aspect ratio, $\lambda = \omega_z/\omega_{\perp} = 20$ and the characteristic oscillator length, $l=2.41\mu m$. The $s$-wave scattering lengths for channels of total spin 0 and 2 are  $a_0 = 101.8  a_B$ and $a_2 = 100.4 a_B$, respectively, where $a_B$ is the Bohr radius. Our simulations run from a spatial extent of $-20 l$ to $20 l$ in both $x$ and $y$ directions with $2001 \times 2001$ grid points. The employed spatial discretization refers to $\Delta x = \Delta y = 0.02 l$, with a time step $\delta t = (2 \times 10^{-4})/\omega_{\perp}$. 

The interplay of the SOC and the external magnetic field yields a distinct arrangement of stable spin configurations (at the saddle points) described by the spin texture \cite{Takeshi_coreless_2004, Kasamatsu_spin_2005}, 
\begin{equation}
\label{eqn:4}
    S_{\alpha} = {\sum\limits_{m_F,n_F=-1}^{1} \psi^*_{m_F} (f_{\alpha})_{m_F,n_F} \psi_{n_F}/\sum\limits_{m_F} |\psi_{m_F}|^2} \quad (\alpha=x,y,z),
    \end{equation}
and the topological charge density, given as 
\begin{equation}
\label{eqn:5}
    q(\mathbf{r}) = \frac{1}{4\pi} \mathbf{s} \cdot \left(\frac{\partial \mathbf{s}}{\partial x} \times \frac{\partial \mathbf{s}}{\partial y} \right)
\end{equation}
Here $\mathbf{s}= \mathbf{S}/|\mathbf{S}|$, with $\mathbf{S} = (S_x, S_y, S_z)$. The associated topological charge $Q$ is defined as $Q =\int q(\mathbf{r}) dx dy$.

\section{Results}
\label{sec:Results}
\begin{figure}[t]

    \centering
    \includegraphics[scale=1,width=0.6\linewidth]{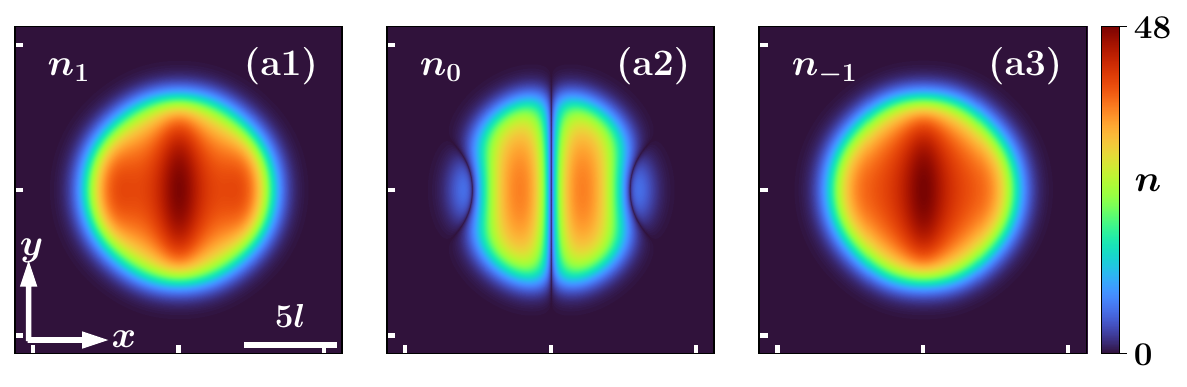}
    \caption{(a1-a3) shows the ground state density profiles $n_1$, $n_0$ and $n_{-1}$ of $m_F=1,0,-1$ components respectively of a ferromagnetic spin-1 BEC of $10^4$ $^{87}\mathrm{Rb}$ atoms in a sinusoidally varying magnetic field $B(\mathbf{r})=(B_0 \sin{x}, 0)$ with strength $B_0 = 0.25 \hbar \omega_{\perp}/ g_F \mu_B l$, in the absence of SOC. The scattering length $a_{0} = 101.8 a_B$ and $a_2 = 100.4 a_B$. The spin-1 BEC is confined in an axisymmetric trap with $(\omega_{\perp}, \omega_z) = 2\pi \times (20,400) \mathrm{Hz}$. The units of length is $\sqrt{\hbar/M\omega_{\perp}}$.}
    \label{fig:1}
\end{figure}

We first numerically solve the coupled GP eqns \ref{eqn:2}-\ref{eqn:3} in imaginary time to generate the ground state solutions of the spinor condensate without restricting the magnetization $m_z$. The ground state densities of the spin-1 BEC, under the influence of the sinusoidally varying magnetic field, in the absence of SOC, are presented in Fig. \ref{fig:1}. The $x$-polarised sinusoidal magnetic field affects the spin-exchange interaction as confirmed from the GP eqns \ref{eqn:2}-\ref{eqn:3}. 

\begin{figure}[b]
    \centering
    \includegraphics[scale=1,width=0.6\linewidth]{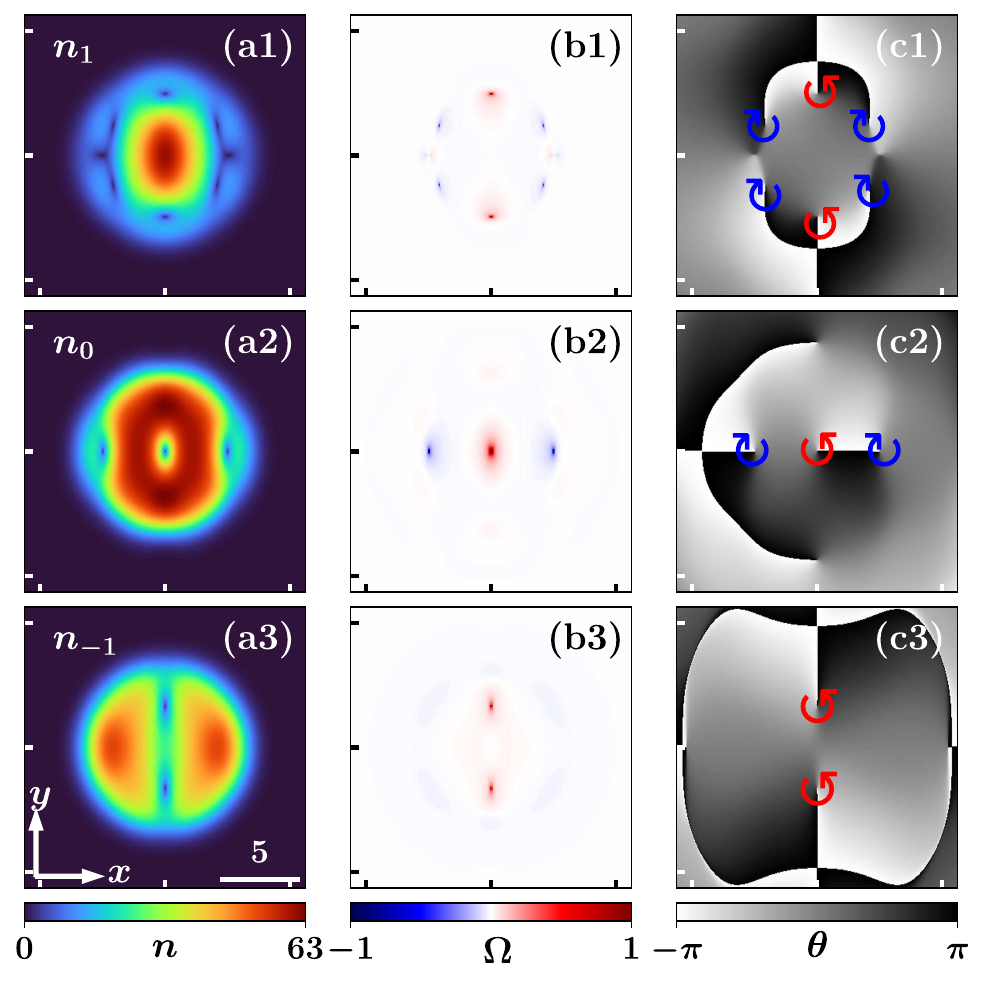}
    \caption{(a1-a3) shows the ground state density profiles $n$ of a SO-coupled ferromagnetic spin-1 BEC of $10^4$ $^{87}\mathrm{Rb}$ atoms in a sinusoidally varying magnetic field $B(\mathbf{r})=(B_0 \sin{x}, 0)$. The SOC strength $\gamma = 0.5 \hbar \omega_{\perp} l$, $B_0 = 0.25 \hbar \omega_{\perp}/ g_F \mu_B l$. The scattering length $a_{0} = 101.8 a_B$ and $a_2 = 100.4 a_B$. (b1-b3) shows the corresponding vorticity profiles which can be measured as $\Omega_{m_F} = \nabla \times J_{m_F}$, with the probability current density $J_{m_F}= \frac{{i}\hbar}{2 M} (\psi_{m_F} \nabla \psi^*_{m_F} - \psi^*_{m_F} \nabla \psi_{m_F})$. c1-c3) represents the corresponding phase profiles $\theta_{m_F} = arg(\psi_{m_F})$. The densities $n_{m_F}$ expressed in units of $l^{-2}$ where $l=2.41 \mu \rm m$ is the characteristic length. The vorticity $\Omega_{m_F}$ is expressed in dimensionless units.} 
    \label{fig:2}
\end{figure} 
In the absence of SOC, the system's symmetry is preserved, and the overall angular momentum remains conserved. However, the sinusoidal magnetic field renders fragments in the density profile of the $m_F=0$ component; although, without any topological excitations in the system. As the saddle points of the sinusoidal magnetic field occur at multiples of $\pi$, the fragment lines pass through $x=0, \pm \pi$. The $m_F = \pm1$ components have a region of relatively higher density located about the $x=0$ line. Without SOC, the system has continuous $SO(2)$ rotational symmetry in the $x$-$y$ plane. However, with Rashba SOC, spin and momentum become coupled, which breaks this symmetry, inducing anisotropy which leads to interesting spin textures.

\begin{figure}[t]
    \centering
    \includegraphics[scale=1,width=0.6\linewidth]{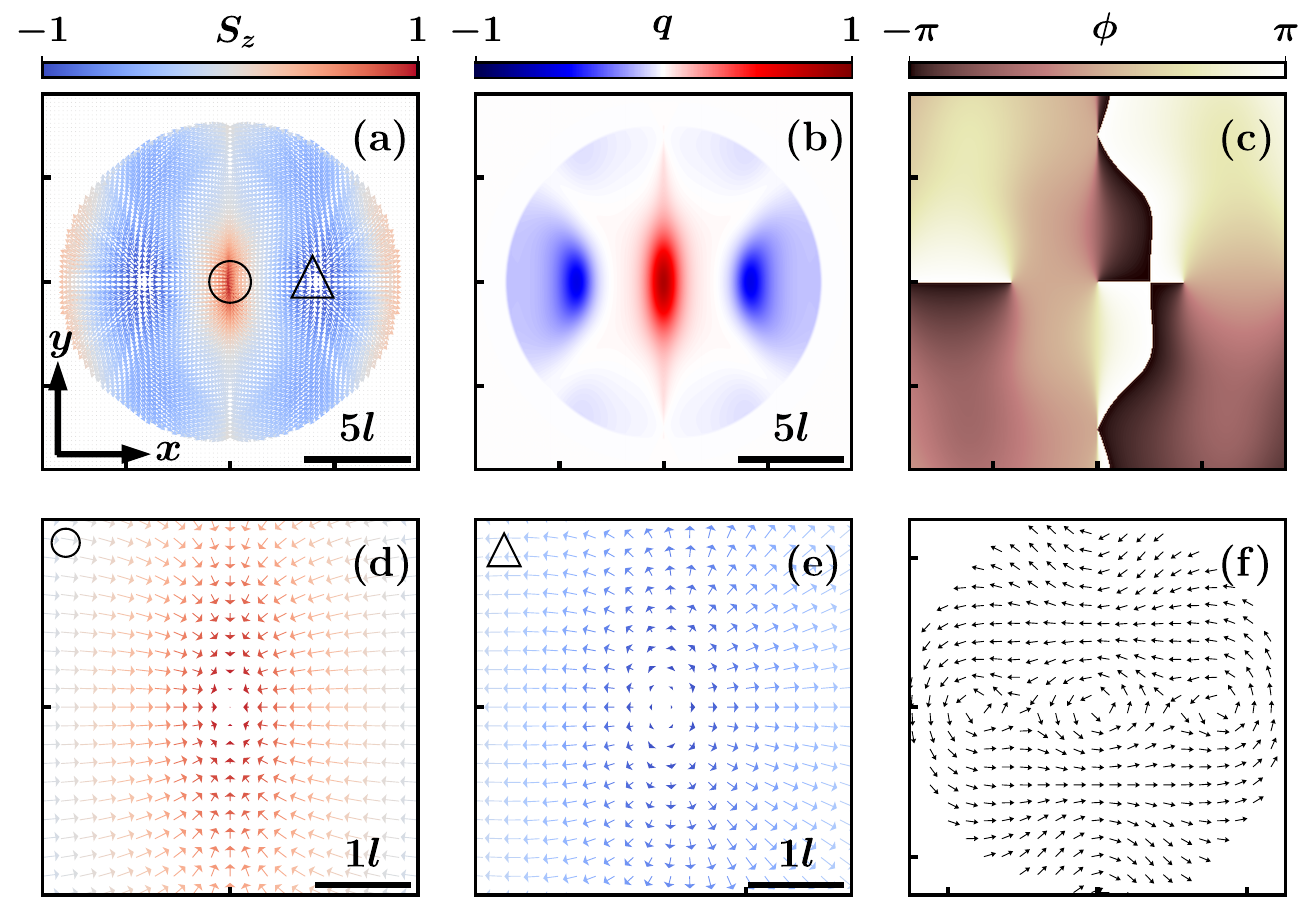}
    \caption{(a) shows the spin texture $\mathbf{S}$ of the initial state demonstrated in Fig. \ref{fig:2}. The arrows represent the transverse $(S_x, S_y)$ components, while the color of the arrows represents the magnitude of $S_z$. The skyrmion and anti-skyrmion are marked by a black circle and a black triangle, respectively. (b) shows the topological charge density $q$ of the equilibrium spin texture. (d) and (e) respectively, are the magnified view of the spin texture of skyrmion and anti-skyrmion shown in (a). (c) shows the transverse spin-phase $\phi$ for the initial state and (f) shows the transverse spin velocity field, \textit{i.e.,} $\vec{\nabla}\phi / |\vec{\nabla}\phi|$.}
    \label{fig:3}
\end{figure}
The combined effect of the external magnetic field and spin-orbit coupling results in density modulations in the system, materializing topological excitations in the form of vortices. In the ground state, we report the formation of a vortex chain (at the saddle points of the magnetic field) in the $m_F=0$ component density [see Fig. \ref{fig:2} (a2)]. The $m_F=1$ component has a local high density at the center [see Fig. \ref{fig:2} (a1)], which is elongated along the $y$-axis with adjacent vortex-antivortex chains surrounding the local high density. The $m_F=-1$ component has vortices along the $y$-axis, with high-density lobes surrounding the axial line [see Fig. \ref{fig:2} (a3)]. The vorticity profile [see Fig. \ref{fig:2} (b1-b3)] clearly shows that the $m_F=0$ component exhibits an alternating vortex-anti-vortex chain. Due to the induced anisotropy, the vortices have a slightly deformed elliptic core. The red (anti-clockwise) and blue (clockwise) arrows in the phase profiles [see Fig. \ref{fig:2} (c1-c3)] mark the circulation quanta of vortices and anti-vortices in the spin-1 BEC.

In the spin texture of the ground state, we confirm a unique anti-skyrmion-skyrmion-anti-skyrmion chain as demonstrated in Fig. \ref{fig:3}(a), where we characterize skyrmions by a stable converging spin pattern with $+ve$ topological charge and anti-skyrmions by a diverging spin pattern with $-ve$ charge. In Fig. \ref{fig:3}(b), we show the topological charge density. The local amplification of the skyrmion and anti-skyrmion structures are presented in Fig. \ref{fig:3}(d) and  Fig. \ref{fig:3}(e), respectively.  The spin phase $\phi = \arctan({S_y/S_x})$ is depicted in Fig. \ref{fig:3}(c). The singularities in the spin phase denote the skyrmions in the guise of spin vortices. Figure \ref{fig:3}(f) gives their associated spin velocity field.

\subsection{ Spin texture dynamics of the SO-coupled spin-1 BEC}
\label{subsec:3.1}
The time evolution of this prepared ground state reveals steady-state dynamics, maintaining a constant total orbital angular momentum, $ L_z^{tot}$ for the spin-orbit coupled Hamiltonian, expressed as \cite{Qu_angular_2018, li2024quantumdropletsmagneticvortices},
\begin{equation}
    L_z^{tot} = x(p_y + \gamma f_x) - y(p_x - \gamma f_y) = L_z^c + L_z^s,
    \label{eqn:6}
\end{equation}
where $L_z^c = xp_y -yp_x$ is the canonical contribution, while $L_z^s= \gamma (x f_x + y f_y)$ is the spin-dependent term. Additionally, the number of particles in each component remains the same throughout the dynamics, thereby preserving the longitudinal magnetization $m_z$. As a matter of fact, the total angular momentum, $L_z^{tot}+m_z$ (orbital plus spin) is a conserved quantity for the steady state [see \ref{apx:a} for discussions]. Building on our investigation of the ground state's steady-state dynamics, we now explore how initializing the system with a predefined longitudinal magnetization $m_z$ influences its static and dynamic properties. 

We now prepare the initial state with fixed magnetization $m_z=0$ \cite{bao_numerical_2008}. The density profile for this state resembles the ground state [Fig. \ref{fig:2}] for which the magnetization settles at $m_z = -0.07$.
To understand the role of magnetization in spin texture dynamics, we then study the real-time evolution of the obtained solutions under two scenarios of spin-mixing dynamics: (a) with freely evolving magnetization by lifting the magnetization constraint, and (b) with fixed magnetization $m_z$ which we delve into in the ensuing section. The fixed magnetization in (b) is achieved by extending the particle normalization to the real-time evolution.

\subsubsection{Case-\rom{1} - Free magnetization dynamics:} 
\label{subsubsec:free}
\begin{figure}[t]
    \centering
    \includegraphics[scale=1,width=\linewidth]{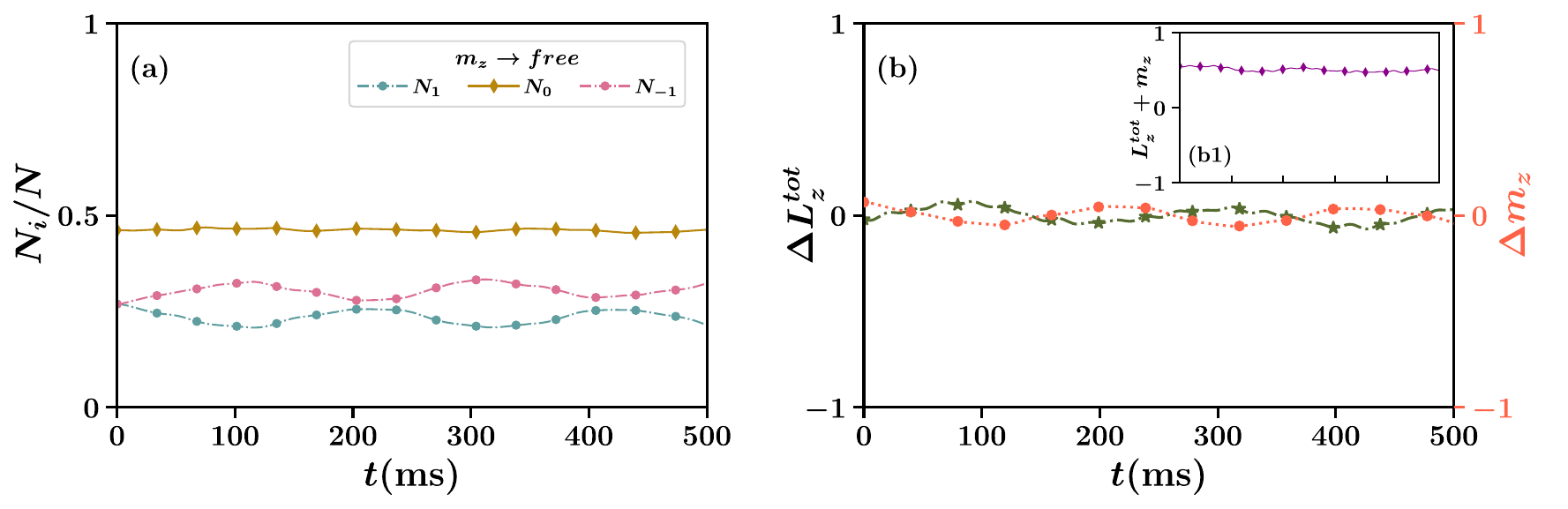}
    \caption{(a) shows the component-wise particle number variation for the cases of skyrmion dynamics with freely evolving magnetization; (b) represents the deviation of orbital angular momentum $\ L_z^{tot}$ and spin angular momentum $m_z$ from their average values and the inset (b1) shows the total angular momentum for the free evolution.}
    \label{fig:4}
\end{figure}
After initializing the system with $m_z = 0$, we allow it to evolve freely. As shown in Fig. \ref{fig:4}(a), the population of particles ($N_{\pm1}$) for the $m_F = \pm 1$ components, oscillates periodically due to the combined effects of spin-mixing dynamics and the applied external magnetic field. This oscillation results in a periodic modulation of $m_z$, highlighting the Einstein–de Haas effect \cite{li2024quantumdropletsmagneticvortices, kawaguchi_einsetin_2006, Qu_angular_2018}. This effect illustrates how the alignment of spins mechanically affects the condensate, further driving the intricate interplay between spin and orbital angular momentum. Consequently, the system exhibits a characteristic exchange \footnote{We calculate the deviation $\Delta A = A - A_{avg}$ for any measurable A. Here, we discuss the variation of $\Delta L_z^{tot}$ vs.\ $\Delta m_z$.} between the spin $m_z$ and total orbital angular momentum $L_z^{tot}$ described in eqn \ref{eqn:6}, as illustrated in Fig. \ref{fig:4}(b), with the total angular momentum depicted in the inset, Fig. \ref{fig:4}(b1). Interestingly, the skyrmion structures maintain relatively stable configurations, experiencing only minimal angular oscillations throughout the relaxed dynamics—a dynamic sequence provided in full in the supplementary materials.

\subsubsection{Case-\rom{2}- Fixed magnetization dynamics:} 
\label{subsubsec:fixed}

\begin{figure}[b]
    \centering
    \includegraphics[scale=1,width=\linewidth]{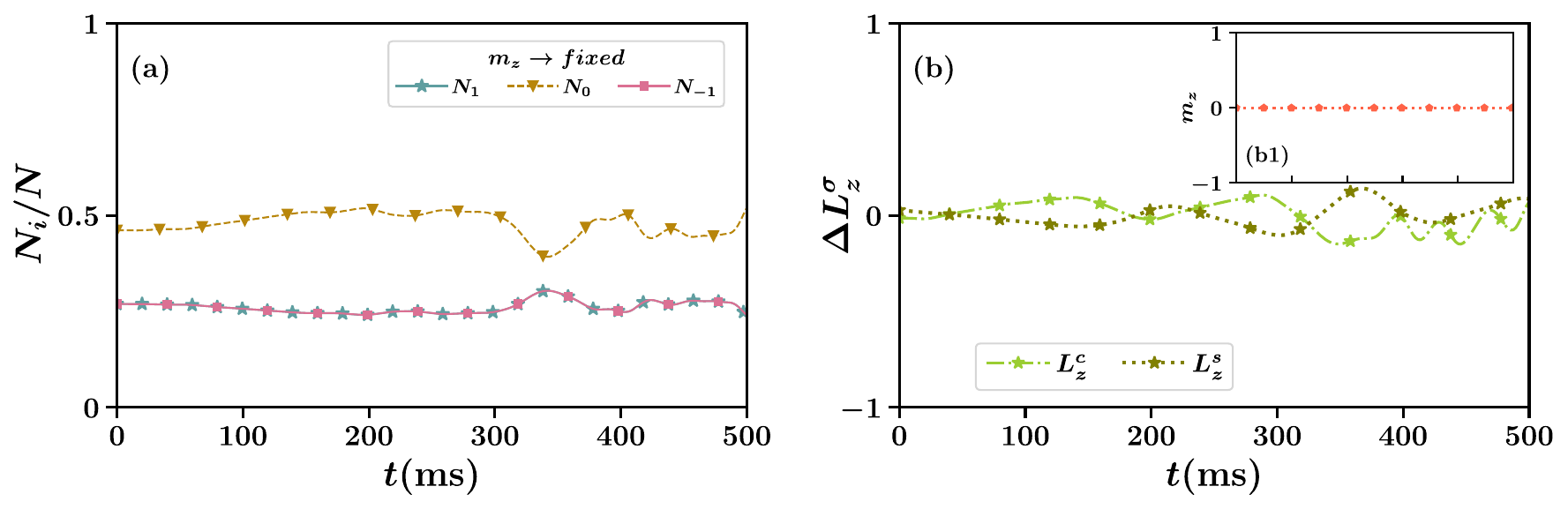}
    \caption{(a) shows the component-wise particle number variation for the cases of skyrmion dynamics with fixed longitudinal magnetization; (b) shows the variation of the canonical and the spin-dependent term in the orbital angular momentum and the inset (b1) shows the $m_z$ variation for fixed magnetization dynamics.}
    \label{fig:5}
\end{figure}
Here, we keep $m_z$ fixed even during the dynamics. With $m_z = 0$, the canonical and spin-dependent angular momenta interact, driving the angular oscillation of the skyrmion chain. Although the magnetization is fixed [see inset Fig. \ref{fig:5}(b1)], spin-exchange collisions enable particle redistribution within the order parameter, as seen in the component-wise population dynamics in Fig. \ref{fig:5}(a). The variations in the canonical contribution $L_z^c$ and the spin-dependent angular momenta $L_z^s$ complement each other, suggesting that the total orbital angular momentum $L_z^{tot}$ still remains nearly constant despite the exchange interactions [Fig. \ref{fig:5}(b)].

\begin{figure}[t]
    \centering
    \includegraphics[scale=1.5,width=\linewidth]{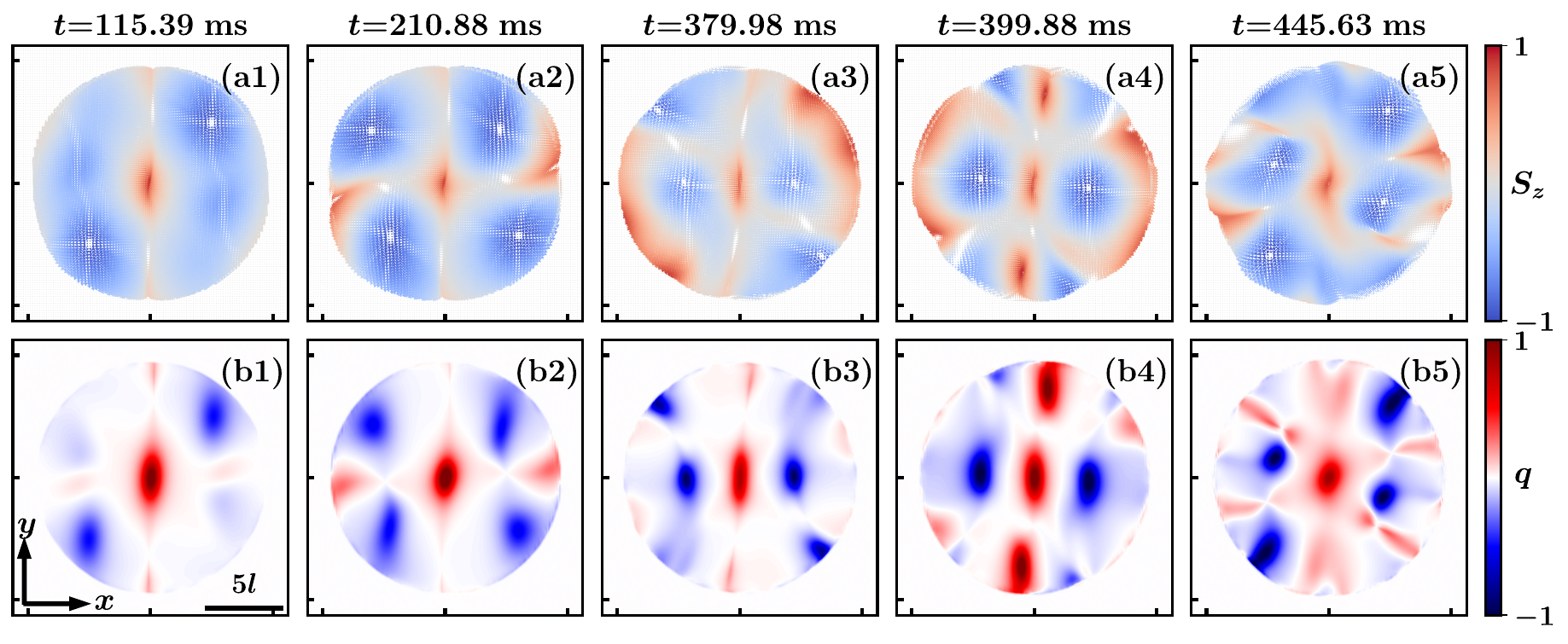}
   \caption{(a1-a5) shows the evolution of the spin texture $\mathbf{S}$ during the dynamics of the SO-coupled spin$\rm{-}1$ BEC with magnetization fixed at $m_z=0$. The color of the arrows represents the magnitude of $S_z$. (b1-b5) shows the topological charge densities associated with these spin textures, respectively, at different time instances shown in columns: (1) $t = 115.39\ \mathrm{ms} $, (2) $t = 210.88\ \mathrm{ms} $, (3) $t= 379.98\ \mathrm{ms} $, (4) $t= 399.88\ \mathrm{ms}$, and (5) $t=445.63\ \mathrm{ms}$. All the parameters are the same as the initial state shown in Fig. \ref{fig:1}. }
    \label{fig:6}
\end{figure}

The constrained evolution of the spin texture and the associated topological charge density (Fig. \ref{fig:6}) drives the oscillatory motion of the skyrmion chain around its initial equilibrium position [similar to Fig. \ref{fig:6}(a)], mimicking the behavior of a simple harmonic oscillator. The anti-skyrmions in the chain move to their extreme positions along the diagonal, return to equilibrium, and then shift to the opposite extreme. Over time, both the spin texture and topological charge shift diagonally from the initial position at the $y=0$ line, reaching an extreme at $t = 115.39\ \mathrm{ms} $ [Fig. \ref{fig:6}(a1, b1)]. From this position, restoring forces from newly formed skyrmions and anti-skyrmions push the chain back to equilibrium at 
$t = 210.88\ \mathrm{ms} $ [Fig. \ref{fig:6}(a2, b2)], reaching it again at 
$t = 379.98\ \mathrm{ms} $ [Fig. \ref{fig:6}(a3, b3)].

As the chain moves toward the opposite diagonal, temporary skyrmions and anti-skyrmions form and disappear, modifying the spin texture. By $t = 399.88\ \mathrm{ms} $, new skyrmions attract the anti-skyrmions to the other side [Fig. \ref{fig:6}(a4, b4)], helping them reach the extreme position at $t = 445.63\ \mathrm{ms} $ [Fig. \ref{fig:6}(a5, b5)]. The system continues to oscillate as it attempts to restore equilibrium, but chaotic dynamics eventually disrupt the spin texture evolution. Throughout, vortices and antivortices appear temporarily, and the oscillatory behavior of the skyrmion chain is analogous to the dynamics of the linear vortex chain in the spin-1 BEC.  A full movie of dynamic evolution is presented in the supplementary material. 

\subsection{Scissors mode excitations in spin texture}
\label{subsection:scissors}
In addition to investigating the system's dynamics under Case-\rom{1} (section \ref{subsubsec:free}) and Case-\rom{2} (section \ref{subsubsec:fixed}), we also analyze several key properties to gain a deeper understanding of the dynamical response of the spinor condensate under a sinusoidally varying magnetic field.
Rashba SOC explicitly breaks the rotational symmetry in the $x$-$y$ plane by coupling spin components to the condensate’s momentum. In the absence of SOC, the system maintains continuous $SO(2)$ rotational symmetry, meaning that collective modes such as the scissors mode follow conventional hydrodynamic behavior in the particle density. However, with Rashba SOC, the canonical momenta are modified as $p_x \to p_x - \gamma f_y$ and $p_y \to p_y + \gamma f_x$. As a result, the superfluid velocity acquires an additional spin-dependent term, $\frac{\gamma}{M} (f_x \hat{y} - f_y \hat{x})$, which introduces anisotropy in the condensate flow as well as in the evolution of the spins \cite{Qu_angular_2018, Qu_scissors_2023}. This means that motion along $x$ and $y$ directions is no longer equivalent, leading to a direct modification of angular dynamics and, consequently, the scissors mode oscillations.

In addition to SOC, the external magnetic field $B(\mathbf{r}) = (\tilde{B_0} \sin x, 0)$ further influences these dynamics by inducing a spatially dependent Zeeman interaction of the form $H_B = \tilde{B_0} \sin x f_x$, where $\tilde{B_0}$ is the strength of the magnetic field. This interaction, however, does not directly modify the superfluid velocity but instead introduces a torque that affects spin precession \cite{kawaguchi_spinor_2012}. The resulting spin evolution equation for $S_z$ is derived from the continuity equation \cite{kawaguchi_spinor_2012},

\begin{equation}
    \frac{\partial S_z}{\partial t} + \nabla \cdot (\mathbf{ v} S_z) = \mathcal{T}_z.
    \label{eqn:7}
\end{equation}
where the torque $T_z$ has two contributions: one due to the external magnetic field,

\begin{equation}
    \mathcal{T}_z^{(B)} = \frac{g_F \mu_B \tilde{B_0}}{\hbar} \sin x S_y.
    \label{eqn:8}
\end{equation}
and another from SOC;

\begin{equation}
    \mathcal{T}_z^{(\rm{SOC})} = \frac{\gamma}{\hbar}(p_x S_x + p_y S_y),
    \label{eqn:9}
\end{equation}

The SOC-induced anisotropy in the superfluid velocity breaks rotational symmetry, leading to angular motion, while the spatially varying Zeeman interaction generates a position-dependent spin torque. We discuss in detail, the evolution of the scissors mode in \ref{apx:c}. Additionally, transient skyrmions introduce further variations in the oscillation structure. These effects are more pronounced in the fixed magnetization case, where the spin texture undergoes enhanced fluctuations compared to the freely evolving system.

\begin{figure}[t]
    \centering
    \includegraphics[scale=1,width=\linewidth]{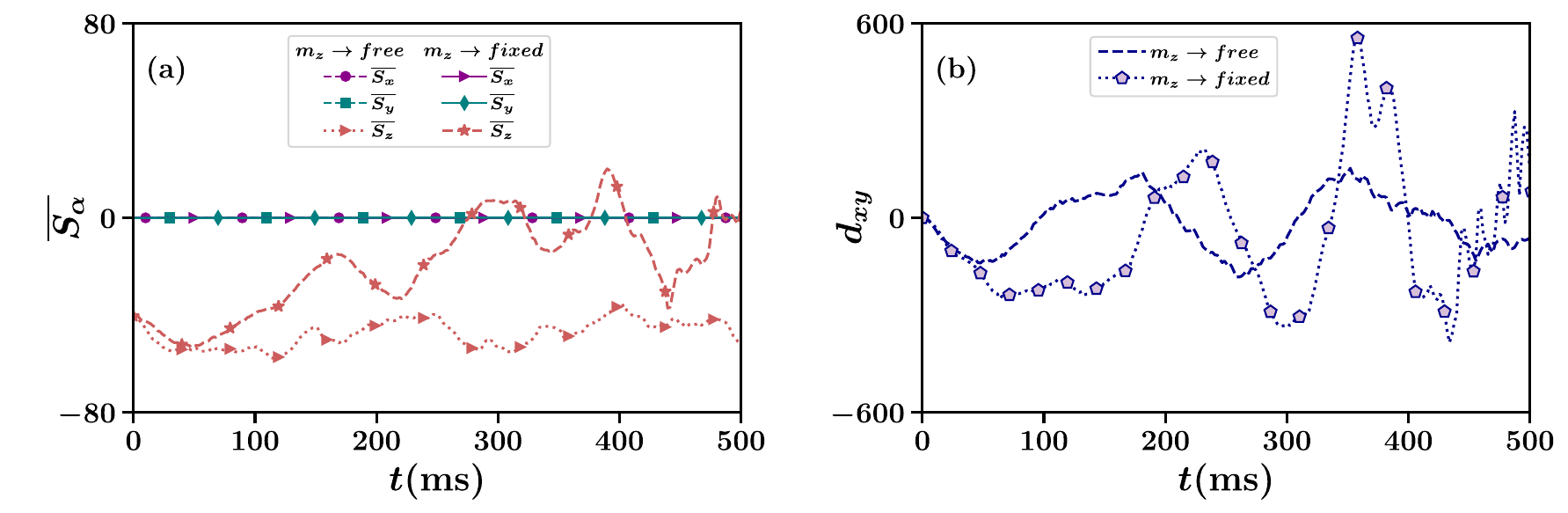}
    \caption{(a) represents the variation of the spatial average $\overline{ S_{\alpha}} = \int S_{\alpha} dx dy$ with time; (b) denotes the scissors modes excitations in the spin texture for both cases.}
    \label{fig:7}
\end{figure}

 In Fig. \ref{fig:7}(a), we show the temporal evolution of the spatial average of the spin components $\overline{S_{\alpha}}$, where $\alpha = (x,y,z)$, to illustrate the spin alignment and fluctuations within the condensate. The spatial average of the transverse spin components $\overline{S_x}$ and $\overline{S_y}$ remain constant at zero over time, indicating no fluctuations in these directions. However, the spatial average of the longitudinal component $\overline{S_z}$ exhibits dynamic variations due to symmetry breaking in the system. In Case-\rom{1}, $\overline{S_z}$ fluctuates locally around a mean value, corresponding to the small amplitude oscillations of the skyrmion chain. For Case-\rom{2}, $\overline{S_z}$ shows more pronounced oscillatory behavior, aligning with the large angular oscillations of the skyrmion chain. This involves creating and annihilating temporary spin structures during the dynamics, contributing to the overall spin texture.  In both cases, we examine the scissors mode being excited in the spin texture to validate the spin fluctuations during the oscillations \cite{Qu_scissors_2023}. As the combined effect of magnetic field and SOC yields a linear skyrmion chain in the $x$-direction, inducing an anisotropy in the spin texture, the scissors mode get excited in the spin texture. We compute $d_{xy} = \langle xyS_z \rangle =\int d^2r \, xy S_z$ as a function of time and confirm the oscillatory response for both cases as presented in Fig. \ref{fig:7}(b). For the free $m_z$ case, the scissors mode oscillates uniformly, but when the system is constrained with longitudinal magnetization $m_z = 0$, the oscillations have non-uniform amplitudes. 

\subsection{Spin-Spin Correlation}
\label{subsection:correlation}

\begin{figure}[t]
    \centering
    \includegraphics[scale=1,width=\linewidth]{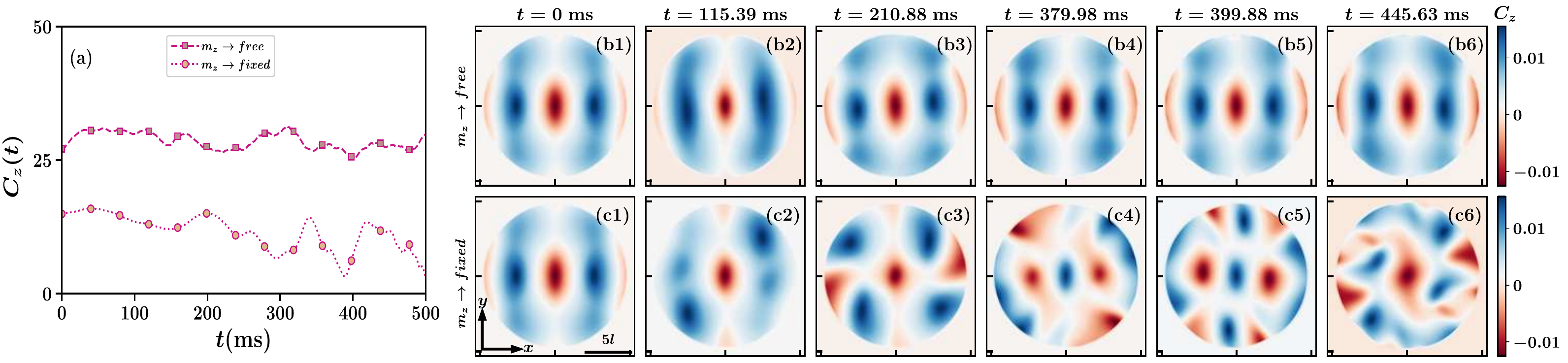}
    \caption{(a) represents the spin-spin time correlations for both cases, (b1-b6) demonstrates the spin-spin space correlations in $S_z$ for various time instances during the dynamical evolution of the system under the free $m_z$ case. (c1-c6) represent the same for the fixed $m_z$ case. All parameters are the same as for the initial state shown in Fig. 1 of the main text. }
    \label{fig:8}
\end{figure}

Further, we examine the spin-spin time correlation for the $z$-component of the spin texture during the dynamics under both cases of magnetization, given as,
\begin{equation}
        C_z(t)= \int \int dt^{\prime} d\mathbf{r} S_z(\mathbf{r},t+t^{\prime})S_z(\mathbf{r},t^{\prime}),
    \label{eqn:10}
\end{equation}
for the dynamics of the initial state prepared with certain fixed $m_z$. We present the variation of $C_z(t)$ w.r.t time in Fig. \ref{fig:8}(a). For Case-\rom{1}, as the system is not constrained, the time correlation owing to the relaxed dynamics has a supposedly long correlation time. The $C_z(t)$ varies marginally over time, \textit{i.e.,} the spin texture is well time-correlated for this case owing to the bounded changes we observe in the spin texture. 
However, for Case-\rom{2}, following the skyrmion chain oscillations, the spin-spin time correlation has an oscillatory response, and as the spin texture gets non-uniform and disturbed over time, the amplitudes of $C_z(t)$ is damped over time.
We analyze the spin-spin space correlation $C_z(\mathbf{r})$ at any time instant, $t$, defined as \cite{saito_kibble_2007,williamson_universal_2016,jose_patterns_2023, huh_universality_2024},
\begin{equation}
    C_z(\mathbf{r},t)= \int dr^{\prime} S_z(\mathbf{r+r^{\prime}},t)S_z(\mathbf{r},t).
    \label{eqn:11}
\end{equation}

We present the space correlations for both cases at various time instances in Fig. \ref{fig:8}. The top panel [Fig. \ref{fig:8} (b1-b6)] shows the $C_z(\mathbf{r},t)$ distributions for the free magnetization case,  and the bottom panel [Fig. \ref{fig:8} (c1-c6)] represents the same for the constrained evolution. At the initial time, $t = 0\ \mathrm{ms}$, the correlation distribution mimics a sinusoidal function along the $x$-axis. During the dynamics, the distribution depicts the dynamical response of the spin texture. In the case of freely evolving magnetization, the system remains very strongly correlated over time owing to the marginal changes in the spin texture. However, the constrained magnetization case reveals an oscillatory response leading to the formation of certain temporary domains in the distribution. These domains also affirm the spin vortices and their evolution with time.

\section{Conclusion}
\label{sec:conclusion}
In conclusion, we have investigated the unique spin texture dynamics of a spin-orbit-coupled spin-1 BEC under a polarized sinusoidally varying magnetic field. In its ground state, the SO-coupled BEC exhibits a linear chain of topologically protected skyrmions at the saddle points of the magnetic field. The dynamical evolution of the system, initiated with $m_z = 0$ magnetization, has been studied under both freely evolving and fixed magnetization conditions to understand the role of magnetization in the spin texture dynamics. The system experiences the Einstein–de Haas effect during the free dynamical evolution, owing to which there is an exchange between the spin and total orbital momenta. Our numerical simulation shows angular oscillations of the skyrmion chain about the initial configuration (akin to spin Bloch oscillations \cite{gangardt_bloch_2009, Kartashov_bloch_2016}) for the fixed magnetization evolution. To further investigate the effect of magnetization on the spin texture dynamics, we provide a brief discussion in the \ref{apx:a}, presenting a comparative study of dynamics with different magnetization values for \textit{e.g.} $m_z = \pm 0.1$. Our preliminary study indicates that increasing SOC strength can produce longer linear skyrmion chains, emphasizing the scalability of the system \cite{saboo_sinusoidal_2024}. Further, relevant studies in classical systems \cite{du2015edge, hou_creation_2018} and ferrofluids \cite{koraltan_generation_2023} suggest controlled skyrmion formation depending on the strength and periodicity of the magnetic field. Such studies can potentially reveal new insights into skyrmion dynamics and quantum control of ultracold gases with spin degrees of freedom. Spinor SO-coupled BECs thus present a promising platform to study quantum fluid properties with solid-state parallels, opening new directions for research in complex spinor systems. This work lays a foundation for future exploration into multi-component spinor BECs, especially under varying the SOC strengths and magnetic field parameters.

\ack
We thank Jae-yoon Choi and Se Kwon Kim for the insightful and enriching email exchanges. AS thanks Subrata Das for his technical assistance and Hari Sadhan Ghosh for his suggestions. We acknowledge the National Supercomputing Mission (NSM) for providing the computational resources of PARAM Shakti at IIT Kharagpur, implemented by C-DAC and supported by the Ministry of Electronics and Information Technology (MeitY) and the Department of Science and Technology (DST), Government of India. A.S. gratefully acknowledges support from the Prime Minister’s Research Fellowship (PMRF), India. S.H. acknowledges the MHRD, Govt. of India, for the research fellowship.

\appendix

\section{Dependence on magnetization values}
\label{apx:a}
\begin{figure}[t]
    \centering
    \includegraphics[scale=1,width=\linewidth]{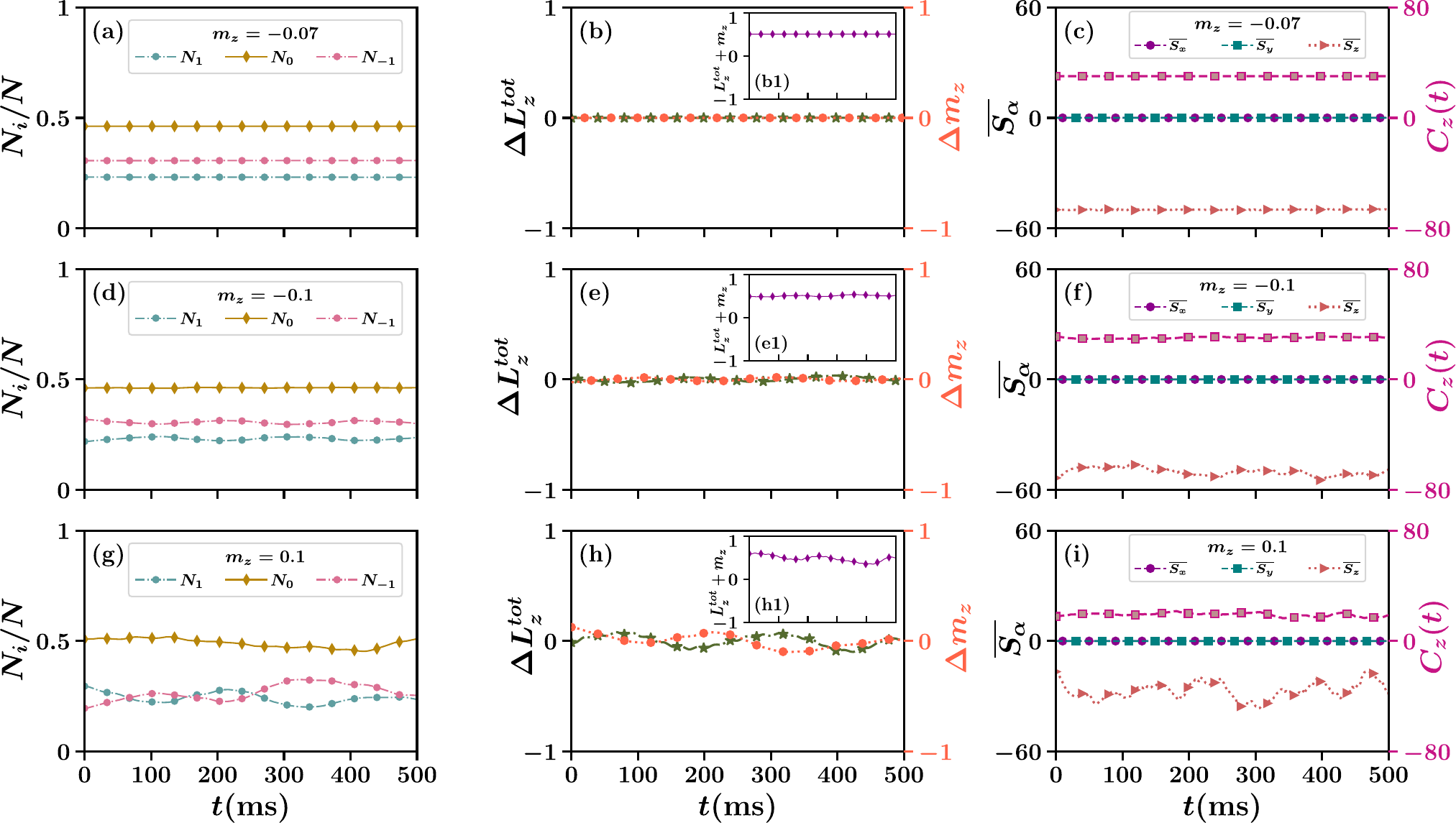}
    \caption{The top panel represents (a) population variation; (b) $\Delta L_z^{tot}$ and $\Delta m_z$ variation with total angular momentum variation in inset (b1); spin texture $S_{\alpha}$ and time correlation $C_z(t)$ variation for the ground state prepared with free magnetization ($m_z=-0.07)$ following a free magnetization dynamics. The middle and bottom panels represent the same for the states prepared with fixed magnetization $m_z = \mp 0.1$, respectively.}
    \label{fig:A1}
\end{figure} 

In this section, we explore different cases for various magnetization values and study the dynamics of the prepared states. From the ground state solutions of the GPEs, the system naturally settles at a magnetization value of $m_z = -0.07$ The dynamical evolution of this state under free magnetization dynamics reveals steady-state behavior. The population variation is presented in Fig. \ref{fig:A1}(a), showing that the particle number in each $m_F$ component remains constant in the steady state. Additionally, both the orbital and spin angular momenta are conserved quantities, as indicated in Fig. \ref{fig:A1}(b). Consequently, the system exhibits the Einstein-de Haas effect [see Fig. \ref{fig:A1} (b1). The steady-state dynamics confirm no significant variation in the spin texture, leading to a constant time correlation [see Fig. \ref{fig:A1} (c)]. 

We further study the dynamics of the system prepared with magnetization $m_z = \mp 0.1$ and draw comparisons from the steady state. As $m_z = -0.1$ is pretty close to the steady state, the dynamic behavior of this state can be analogously understood. During the free magnetization evolution, the system experiences an oscillatory population variation [see Fig. \ref{fig:A1} (d)] due to the allowed particle exchange during the spin-mixing dynamics. This is also reflected in the oscillatory $m_z$ variation, which complements the $L_z$ variation, indicating the possibility of orbital and spin angular momentum exchanges as shown in Fig. \ref{fig:A1} (e). The Einstein-de Haas effect [see Fig. \ref{fig:A1} (e1)] testifies for this angular momentum exchange. As this system is slightly deviated from the steady state, the spin texture, $S_z$ in particular shows slight variation with time and thus its time correlation $C_z$ as shown in Fig. \ref{fig:A1} (f). For $m_z = 0.1$, the dynamical evolution is slightly offset (as the state is far from the steady state), as confirmed by the population variation shown in Fig. \ref{fig:A1} (g). The $L_z$ and $m_z$ variations are more pronounced [see Fig. \ref{fig:A1} (h)], which leads to slight variation in the total angular momentum of the system [inset Fig. \ref{fig:A1} (h1)]. During the free magnetization evolution, the spin texture and the correlation in $S_z$ show significant variations as presented in Fig. \ref{fig:A1} (i). 

Thus, longitudinal magnetization $m_z$ plays a pivotal role in leading the system from a well-correlated state to one with noticeable differences.

\section{Dynamical variation of topological charge}
\label{apx:b}
\begin{figure}[t]
        \centering
        \includegraphics[width=0.75\textwidth]{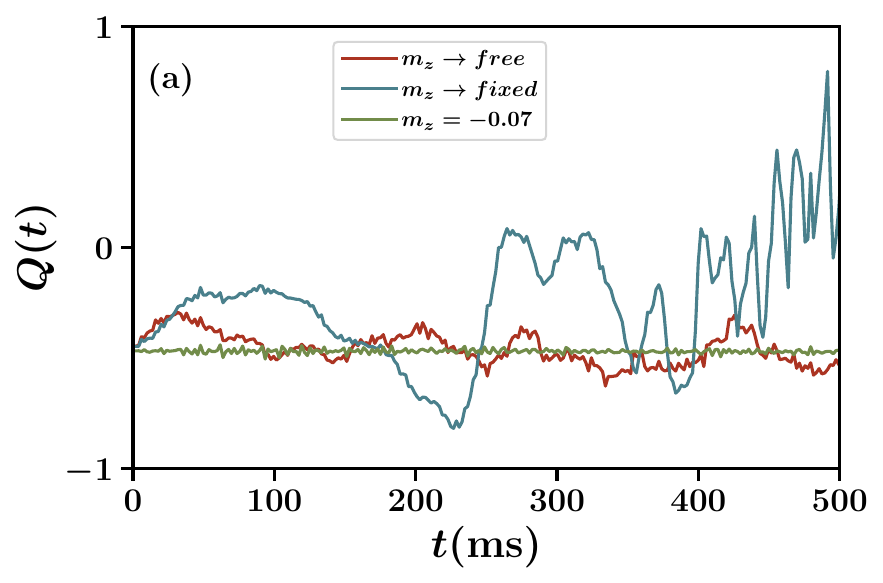}
        \caption{Variation of topological charge Q(t) with time for the fixed and freely evolving magnetization dynamics in comparison to the dynamical evolution of the ground state prepared without any pre-fixed magnetization ($m_z = -0.07$) as a reference.}
        \label{fig:B1}
\end{figure}

We present the variation of the topological charge $Q(t)$ over time in Fig. \ref{fig:B1}. This analysis validates the temporary formation and disappearance of skyrmions in the spin texture shown in Fig. 5 of the main text. Specifically, we compare the fixed and freely evolving cases using the dynamical evolution of the ground state prepared without any pre-fixed magnetization ($m_z = -0.07$) as a reference. While this reference state shows a reasonably constant $Q(t)$, indicating an overall stable spin texture (as discussed in \ref{apx:a}), the cases initialized with $m_z = 0$ exhibit distinct behaviors. For the freely evolving case, $Q(t)$ oscillates due to minimal spin texture variations reflecting the motion of the skyrmions, whereas for the fixed magnetization case, $Q(t)$ fluctuates with greater magnitude due to the motion as well as the formation and annihilation of temporary skyrmions and anti-skyrmions, confirming the transient nature of these topological structures.

\section{Detailed calculation of the scissors mode in spin texture}
\label{apx:c}
To further quantify the effect of SOC and the external magnetic field on the scissors mode in the spin texture, we analyze the moment $\langle xyS_z \rangle$,

\begin{equation}
    d_{xy} =  \langle xy S_z \rangle \equiv \int d^2r \, xy S_z.
\end{equation}
Taking it's time derivative:

\begin{equation}
    \frac{d}{dt} \langle xy S_z \rangle = \int d^2r \, \left(  xy \frac{\partial S_z}{\partial t} \right).
\end{equation}
We substitute \( \frac{\partial S_z}{\partial t} \) from the spin continuity equation \ref{eqn:7}:

\begin{equation}
    \frac{d}{dt} d_{xy} = \int d^2r \, \left(  - xy \nabla \cdot (\mathbf{v} S_z)+ xy \mathcal{T}_z \right).
\end{equation}
Applying integration by parts to the divergence term (assuming boundary terms vanish):

\begin{equation}
    \int d^2r \, xy \nabla \cdot (\mathbf{v} S_z) = -\int d^2r \, (y v_x + x v_y)S_z.
\end{equation}
Therefore, 

\begin{equation}
    \frac{d}{dt} d_{xy} = \int d^2r \, (y v_x + x v_y)S_z + \int d^2r \, xy \mathcal{T}_z.
\end{equation}
Taking another time derivative:

\begin{equation}
    \frac{d^2}{dt^2} d_{xy} = \int d^2r \, \left[  \left(x \frac{\partial (v_y S_z)}{\partial t} + y \frac{\partial (v_x S_z)}{\partial t}\right) + xy \frac{\partial \mathcal{T}_z}{\partial t} \right].
\end{equation}
Now expanding,

\begin{equation} \frac{\partial \mathcal{T}_z}{\partial t}  = \frac{\partial \mathcal{T}_z^{(B)}}{\partial t} + \frac{\partial \mathcal{T}_z^{\rm{(SOC)}}}{\partial t}.\end{equation}
Now,
\begin{equation} \frac{\partial \mathcal{T}_z^{(B)}}{\partial t}  = \frac{\partial }{\partial t} \left(\frac{g_F \mu_B \tilde{B_0}}{\hbar} \sin x S_y\right)=  \frac{g_F \mu_B \tilde{B_0}}{\hbar} \sin x \frac{\partial S_y}{\partial t}
\end{equation}

In our setup, the external field is along the $x$ direction. When a magnetic field is applied, the spin precesses. For small deviations, the spin precession gives a relation;
\begin{equation} 
\frac{\partial S_y}{\partial t}  = -\omega_B \sin x S_z,
\end{equation}
with the Larmor frequency
\begin{equation} 
\omega_B  = \frac{g_F \mu_B \tilde{B_0}}{\hbar},
\end{equation}
Therefore, 

\begin{equation}
\frac{\partial \mathcal{T}_z^{(B)}}{\partial t} = -{\omega_B}^2 \sin^2 x S_z = -{\omega_B}^2(1- \cos^2 x)S_z 
\end{equation} 
Here the minus sign reflects the direction of precession.

Now, let us consider the $\mathcal{T}_z^{\rm{SOC}}$,

\begin{eqnarray} \frac{\partial \mathcal{T}_z^{(\rm{SOC})}}{\partial t}  &= \frac{\gamma }{\hbar}\frac{\partial }{\partial t} \left(p_x S_x + p_y S_y\right) \nonumber \\&=  \frac{\gamma }{\hbar} \left(p_x \frac{\partial S_x}{\partial t} + \frac{\partial p_x}{\partial t}F_x + p_y \frac{\partial S_y}{\partial t} + \frac{\partial p_y}{\partial t} F_y\right)
\end{eqnarray}

In the hydrodynamic (or semiclassical) limit the changes in $p_{x,y}$
  due to external potentials are often subleading compared to the changes in the spin degrees of freedom. Thus, we focus on the terms

\begin{equation}
    \frac{\partial \mathcal{T}_z^{(\rm{SOC})}}{\partial t} =  \frac{\gamma }{\hbar} \left(p_x \frac{\partial S_x}{\partial t} + p_y \frac{\partial S_y}{\partial t}
    \right)
\end{equation}   
For a system with Rashba SOC the Hamiltonian contains a term
\begin{equation}
H_{\rm{SOC}} = -\gamma (p_x f_y - p_y f_x)
\end{equation} 
From the Heisenberg equation of motion,

\begin{equation}
    \frac{d S_\alpha}{dt} = \frac{i}{\hbar} [H_{\rm{SOC}}, S_\alpha],
\end{equation}
we obtain the time evolution of the spin components:

\begin{equation}
    \frac{\partial S_x}{\partial t} \approx \frac{\gamma}{\hbar} p_x S_z, \quad \frac{\partial S_y}{\partial t} \approx - \frac{\gamma}{\hbar} p_y S_z.
\end{equation}
Substituting these into our previous equation for the time derivative of the SOC torque,

\begin{equation}
    \frac{\partial \mathcal{T}_z^{(\rm{SOC})}}{\partial t} = - \frac{\gamma^2 }{\hbar^2} \left(p_y^2 - p_x^2\right)S_z.
\end{equation}
We obtain the final equation for the spin-texture scissors mode:

\begin{eqnarray} 
\frac{d^2}{dt^2} d_{xy} &+ \omega_B^2 d_{xy} - \int d^2r \left(x \frac{\partial (v_y S_z)}{\partial t} + y \frac{\partial (v_x S_z)}{\partial t}\right)  \nonumber \\ & - {\omega_B}^2\int d^2r x y \cos^2 x S_z + \frac{\gamma^2 }{\hbar^2} \int d^2r xy (\partial_x^2 - \partial_y^2)S_z = 0. \end{eqnarray}

This confirms that the scissors mode oscillates in the spin texture.

\bibliographystyle{iopart-num} 
\bibliography{spin}      

\end{document}